\documentclass[twocolumn,prl,aps,showpacs]{revtex4}
\usepackage{epsfig}
\usepackage{fancyhdr}
\usepackage{verbatim}

\begin{document}

\title{Experimental Free Energy Surface Reconstruction From
Single-Molecule Force Spectroscopy Using Jarzynski's Equality}

\author{Nolan C.\ Harris, Yang Song, and Ching-Hwa Kiang$^*$}

\affiliation{Department of Physics and Astronomy, Rice University,
Houston, TX\ \ 77005}

\begin{abstract}
We used the atomic force microscope to manipulate and unfold individual
molecules of the titin I27 domain and reconstructed its free energy
surface using Jarzynski's equality. The free energy surface for both
stretching and unfolding was reconstructed using an exact formula
that relates the nonequilibrium work fluctuations to the molecular free
energy. In addition, the unfolding free energy barrier, {\em i.e.}\ the
activation energy, was directly obtained from experimental data for the
first time.  This work demonstrates that Jarzynski's equality can be
used to analyze nonequilibrium single-molecule experiments,
and to obtain the free energy surfaces for molecular systems,
including interactions for which only nonequilibrium work can be measured.
\end{abstract}



\maketitle
\thispagestyle{fancy}

One way to probe molecular properties is to drive a system out of
equilibrium and to observe the response. Interpretation of data from
dynamic measurements allows one to reconstruct both the equilibrium
properties of molecules and responses to external perturbations
\cite{Intro}.  Equilibrium parameters are
usually deduced from kinetic measurements, and it remains challenging
to relate nonequilibrium distribution data to equilibrium properties
\cite{Fox03a}.  Advances in single-molecule manipulation and
measurement techniques have made it possible to directly probe the
dynamics of molecular interactions
\cite{Fernandez97a,Nelson00a}. The nonequilibrium work
theorem, {\em i.e.}\ Jarzynski's equality \cite{Jarzynski97a}, relates
nonequilibrium measurements of nanoscale systems to equilibrium
properties \cite{Szabo01a,Szabo05a,Jarzynski06a}.  It promises to
extract thermodynamic parameters such as free energies from
single-molecule measurements.

Forced unfolding of single molecules, now achievable using the atomic force
microscope (AFM) and laser optical tweezers, has been used to probe the
molecular interactions and mechanical properties of individual
molecules \cite{Fernandez97a,HansmaP99b}. In these experiments, single
molecules are held at both ends and stretched while the cantilever
spring restoring force ($F_s$) is measured. Applying an external force
drives the system out of equilibrium, and transitions between
states are directly observed as the system settles to a new equilibrium
state. However, interpretation of these results and deduction of
equilibrium properties from these nonequilibrium measurements remains
controversial \cite{Fernandez99a,Clarke03b,Szabo03a,Thirumalai06b}.

\begin{figure}[!b]
\begin{center}
\epsfig{file=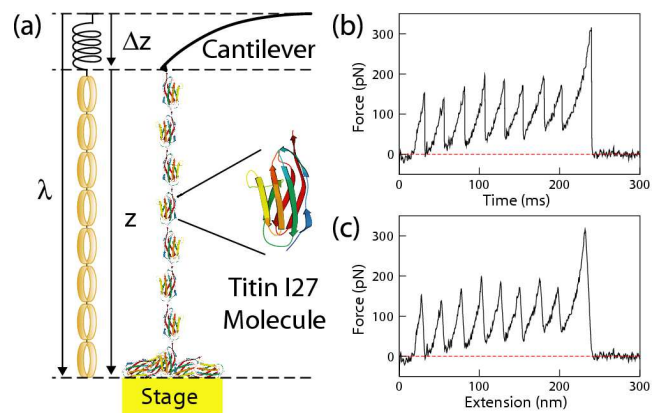,clip=}
\vspace{-0.2in}
\end{center}
\caption{(color).  Single-molecule pulling experiments using AFM.
(a) One end of the molecule is attached to the cantilever tip and the
other end to a gold substrate, whose position is controlled by a
piezoelectric actuator.  An analogue of the
single-molecule force measurements is illustrated. The cantilever
spring obeys Hooke's law, whereas the protein molecular spring follows
the worm-like chain model (illustrated using rubber bands). (b) A
representative force versus time trace, taken at 1.00 $\mu$m/s using a
cantilever with a spring constant of 0.04 N/m. Each force peak
represents unfolding of an individual titin I27 domain, with the final
peak resulting from the detachment of the molecule from the AFM tip.
(c) Corresponding force-extension curve. The tip force baseline 
was determined using the part of the force curve where the
molecule is completely detached from the tip, when the cantilever
spring is at its equilibrium position.  
\label{AFM}}
\end{figure}

It has been widely anticipated that equilibrium free energy differences
can be derived from nonequilibrium measurements using Jarzynski's
equality \cite{Jarzynski97a}. The difference in equilibrium free
energy, $\Delta G$, is related to the fluctuations of work performed
during a nonequilibrium process, $W_\lambda$, by
\cite{Jarzynski97a,Jarzynski06a}
\begin{equation}
\label{jarzynski} 
\langle {e^{-\beta W_\lambda}} \rangle_N \equiv \int
dW_\lambda \rho(W_\lambda) e^{-\beta W_\lambda} = e^{-\beta \Delta G}
\end{equation}
where $\beta=(1/k_B T)$, $k_B$ is the Boltzmann constant, and $T$ is
the temperature of the thermal bath. The $\langle ... \rangle_N $
represents an average over $N$ realizations of the process, and the
equality is exact in the limit $N \rightarrow \infty$. The
nonequilibrium work distribution, $\rho(W_\lambda)$, depends on the
schedule for varying the work parameter $\lambda$, which is the
external perturbation. The equality is simple; however, its application
to interpreting single-molecule results is not straightforward. The
equation involves the thermodynamic work done on the system and the controlled work
parameter with $W_\lambda = \int F \cdot d \lambda$. In AFM
experiments, the system includes the cantilever spring and the molecule
plus water, and $\lambda$ refers to the change in cantilever anchor to
stage distance (see Fig.~\ref{AFM}), not the tip-to-sample distance,
which measures the molecule end-to-end distance $z$, {\em i.e.}\ the
order parameter, or reaction coordinate.

We briefly review the experimental setup to which 
Eq.~(\ref{jarzynski}) applies.
Consider at time $t=0$, the system is at an equilibrium state
$\lambda(0)=\lambda_A$. We perform external work on the system by
controlling the work parameter following a $pre-determined$ schedule,
$\lambda(t)$, from an initial state $\lambda_A$ to a final state
$\lambda_B$. The system is then allowed to relax to equilibrium while
$\lambda$ is held constant at $\lambda_B$. Since we do not perform
external work on the system during relaxation, we can omit this last
step and the equality still holds. Hence Jarzynski's equality
allows us to determine the $G(\lambda)$ from an equilibrated state $A$
to an arbitrary state $B$.

A proof-of-principle experiment and molecular dynamics simulations
testing the Jarzynski estimator have been 
performed \cite{Bustamante02c,Schulten03a}. 
The experimental test involved stretching
individual RNA molecules reversibly and irreversibly using optical
tweezers, and the free energy of unfolding, {\em i.e.}\ the
stability of the molecules, was determined.  
However, the usefulness of Jarzynski's equality lies with its ability 
to obtain directly the entire free energy landscape, which could only
be estimated using kinetic approaches to date \cite{Kinetics}.
We will show that Jarzynski's equality can be used to determine
directly the free energy profile of molecular stretching and 
unfolding, including the free energy barrier of unfolding.

\begin{figure}[!b]
\begin{center}
\epsfig{file=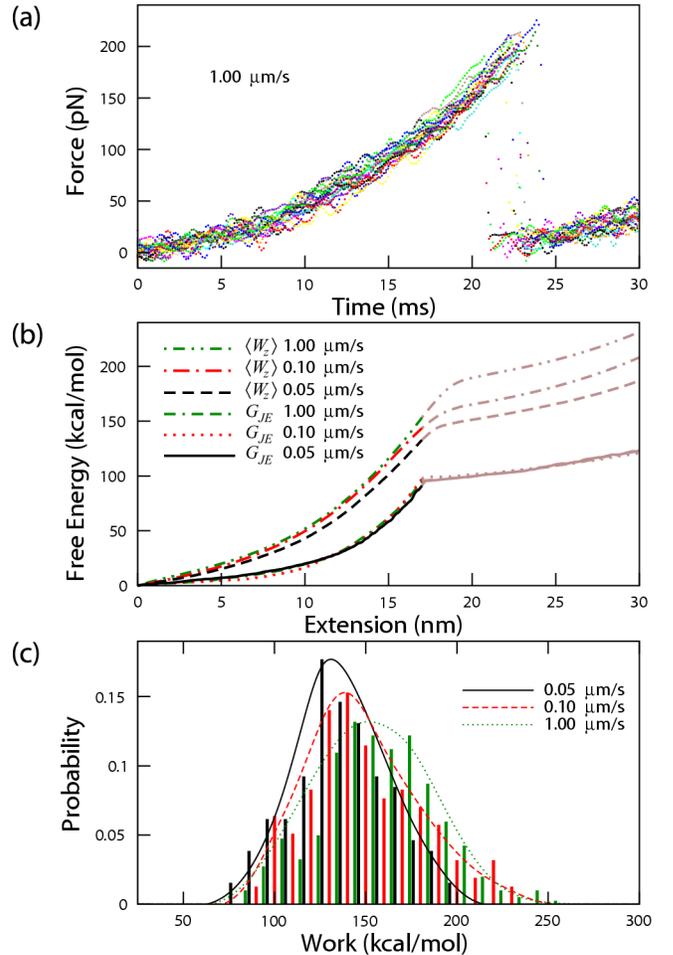,clip=}
\vspace{-0.2in}
\end{center}
\caption{(color).  
Free energy reconstruction of titin I27 for
pulling velocities of 0.05, 0.10, and 1.00 $\mu$m/s obtained using 64,
132, and 226 curves, respectively. 
(a) Typical unfolding force versus time curves for titin I27 
domain taken at 1.00 $\mu$m/s.
Shown are 20 curves smoothed using a smoothing spline for display purposes. 
(b) Free energy $G(z)$ calculated using the Jarzynski estimator, 
$G_{JE}$, applied to the raw data.  The averaged work, 
$\langle W_z \rangle = \sum_i^N W_z /N$, where $W_z = \int F \cdot dz$,
is displayed for comparison.  $\langle W_z \rangle$ is larger than 
the equilibrium 
free energy $G_{JE}$ by about a factor of 2 and is velocity
dependent, whereas $G_{JE}$ is velocity independent. 
The curves are accurate up to the transition state (solid line).
(c) Distributions of work for $z$ as a function 
of pulling velocity. The calculated work includes stretching and unfolding one domain. 
The curve fit to each distribution is a smoothing spline fit to the 
data as a guide to the eye. 
\label{G_curve}}
\end{figure}

Our system of interest is the mechanical unfolding of the I27 domain of
human cardiac titin \cite{WangK79a}. 
The mechanical properties of the immunoglobulin (Ig)-like
domains are directly correlated with the
protein's biological function in the muscles \cite{Fernandez97a}.
The kinetic barrier of these mechanical proteins is important
in determining the dynamic behavior of the proteins during the
stretch-release process. Therefore, the titin free energy surface,
including the unfolding barrier height, is useful for quantification
of titin's function in the heart muscle.

We used AFM to stretch individual molecules of eight serially linked
repeats of the titin I27 domain, as illustrated in Fig.~\ref{AFM}.
The protein was stretched when the substrate stage was moved by
$\lambda$, which was set at a constant velocity $v$, {\em i.e.}\
$\lambda = vt$. 
The cantilever displacement from its equilibrium position $\Delta z$ was
recorded, and the molecular end-to-end distance as a function of time
was calculated using $z = \lambda - \Delta z$.
The force curves are aligned using the best worm-like chain (WLC) 
fit of the force below
the unfolding force.  This method has been shown to minimize the effect of
instrument drift that affects the measured values \cite{Bustamante05d}. 
To correctly calculate $\Delta G$ as a function of the molecular
end-to-end distance, we used an {\em exact} expression that connects
the nonequilibrium fluctuations of work to the Gibbs free energy $G(z)$
\cite{Szabo05a}
\begin{equation}
\label{Gt} 
e^{-\beta G(z)} = \langle \delta(z-z_t) e^{-\beta [W_z(t) -
U_0(z_0,\lambda_A)]} \rangle_N
\end{equation}
where $z_0$ and $z_t$ are the end-to-end distances of the molecule at
times 0 and $t$ during one realization of the process, $F_m$ is the
force on the molecule, $W_z(t)$ is the mechanical work done on the
molecule up to time $t$, $\delta(z-z_t)$ is the Dirac $\delta$
function, and $U_0$ is the potential energy stored in the cantilever
spring at time 0.

To calculate $G(z)$ using Eq.~(\ref{Gt}), we divided each of the $N$
trajectories of duration $\tau$ into discrete time steps $\delta t$ so
$T=\tau/\delta t$, where $T$ is the total number of time steps in a
given trajectory,
\begin{eqnarray}
\label{Gb}
\exp[-\beta G(z^{(m)})] &\approx& {1 \over NT}\sum_{n=1}^N\sum_{s=1}^T
\delta_\epsilon(z^{(m)}-z_{n,s}) \nonumber \\
& & \quad \exp(-\beta[W_{n,s} - U(z_{n,0},\lambda_A)])
\end{eqnarray}
where $z_{n,s}$ is $z$ at the $s$'th time step for the $n$'th
trajectory, $z_{n,0}$ is the initial value of $z$ for the $n$'th
trajectory, and $W_{n,s}$ is the work performed up to time $t_s=s \cdot
\delta t$ for the $n$'th trajectory. We divided the $z$-axis into bins
of width $\epsilon$ and let $z^{(m)}$ represent the mid-point of the $m$'th
bin.   The $\delta$ function is $1/\epsilon$ when $z_{n,t}$ falls
inside the $m$'th bin and $0$ otherwise. The integration starts from
the beginning of the curve, where the cantilever is close to its
resting position, $z=0$, at $t=0$. This initial condition is required
for using Jarzynski estimator, which states that the process needs to
start from an equilibrated state. It is also advantageous when using
Eq.~(\ref{Gb}) that the initial energy stored in the cantilever
spring, $U_0(z_0,\lambda_A)$, is close to $0$. 
We compare this result to the {\em approximate} free energy surface 
derived from 
\begin{equation}
\label{Gz} 
e^{-\beta G_z} \approx \langle e^{-\beta \int F_m \cdot dz - U_0(z_0,\lambda_A)} \rangle_N
\end{equation}
The unfolding free energy surface of titin I27 determined from
Eq.~(\ref{Gz}) and Eq.~(\ref{Gb}) are very similar, perhaps
due to the relatively stiff cantilever used in AFM.  However,
it is physically and theoretically more meaningful to use 
Eq.~(\ref{Gb}), since
determination of the entire free energy surface relies on 
converting the coordinate from $t$ to $z$.
Figure~\ref{G_curve} displays the free energy surface measured
at three different velocities, determined using Eq.~(\ref{Gb}).

The unfolding free energy barrier $\Delta G_u^\ddagger$ 
can be calculated from the reconstructed free
energy curve. Using 0.6~nm as the distance between the native and the
transition state ($x_u$) \cite{Clarke03b,Bustamante04b}, 
we calculated the unfolding
free energy barrier $\Delta G_u^\ddagger$ for pulling velocities of 
0.05, 0.10, and 1.00 $\mu$m/s, to be 
11.0, 11.7, and 11.4 kcal/mol, from Eq.~(\ref{Gb}) and 
11.5, 11.5, and 10.7 kcal/mol from Eq.~(\ref{Gz}), respectively.
The uncertainty in the averaged $\Delta G_u^\ddagger=$ 11.4 and 11.2 
kcal/mol, calculated using the bootstrap method, is 0.4 and 0.3 kcal/mol
for Eqs.~(\ref{Gb}) and (\ref{Gz}), respectively.
This result compares favorably to an estimated value of 10--16 kcal/mol
\cite{Szabo03a,Szabo06a,Clarke04a}.
A major source of error for $\Delta G_u^\ddagger$ from Jarzynski
estimator comes from the uncertainty in $x_u$.
Using the largest estimated error of 0.07~nm uncertainty in 
$x_u$ \cite{Clarke03b}, the estimated uncertainty of $\Delta G_u^\ddagger$
is 1.2 kcal/mol.

The free energy surface is accurately reconstructed from
$z$ = 0 to 17~nm, the transition state.
The free energy of unfolding is insensitive to the distance of
reconstruction.  As an example, if we use 15~nm or 19~nm,
$\Delta G_u^\ddagger$ changes by 2.5 and 0.4\%, respectively, for
the pulling velocity of 0.05 $\mu$m/s.
The vast majority of the proteins in the ensemble are in the folded
state (99.9997 \% using the free energy $\Delta G_u$ from Ref.~\cite{Baker01a})
so the contribution from the initially unfolded proteins is negligible.
To minimize the contribution from other unfolded domains to the measured
free energy, we analyzed only the first domain stretching event.
Using all domain unfolding events in the analysis changes
$\Delta G^\ddagger_u$ by less than 2\%.

\begin{figure}[!b]
\begin{center}
\epsfig{file=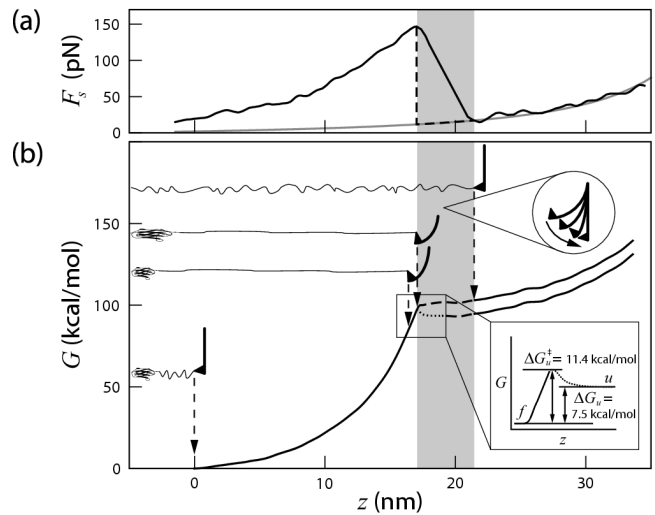,clip=}
\vspace{-0.2in}
\end{center}
\caption{Free energy surface of titin I27.
(a) A typical force versus extension curve.  
The gray curve is the WLC fit to the
following domain.  The shaded region indicates that, when the domain
ruptures and the cantilever snaps, the force on the molecule is not
registered and, therefore, the free energy surface may not be
recovered with high accuracy.  The dashed line indicates an
approximation of the force exerted on the molecule.  
(b) The free energy surface of unfolding titin I27.  The cantilever
position and the molecular extension at each stage are illustrated.
The curve is composed of the reconstructed free energy surface up to
the transition state (solid) and estimated free energy change
\cite{Baker01a} and distance \cite{Plaxco04a} beyond the transition
state (dotted).
\label{stretching}} 
\end{figure}

Note that it is not possible to compare our results directly to 
published values, since $\Delta G_u^\ddagger$ has not been determined,
and only kinetic information is available.  Chemical denaturant studies gave 
an estimated unfolding rate constant, $k_u^0$, of 
6$\times$10$^{-4}$s$^{-1}$ \cite{Clarke04a}, 
while forced-unfolding studies gave an estimated $k_u^0$ of 
10$^{-3}$--10$^{-6}$s$^{-1}$
\cite{Fernandez97a,Fernandez99a,Clarke03b,Szabo03a}, and the $\Delta
G_u^\ddagger$ was then calculated using $k_u^0= k_0 e^{-\beta \Delta
G_u^\ddagger}$. Since the prefactor $k_0$ of protein unfolding is
unknown, the free energy barrier can only be estimated by this procedure
\cite{Eaton02a,Clarke02b}.  However, combining our free energy determination
with the kinetic information, we can determine the prefactor $1/k_0$ to be
6 $\mu$s, which lies within the expected range \cite{Eaton02a,Thirumalai04a}.

The free energy surface immediately past the transition state
cannot be reconstructed with high accuracy from constant velocity
unfolding experimental data.  This is because the force exerted on the molecule
is discontinuous when the domain ruptures and expands against the cantilever. 
In the region where the
domain ruptures and the cantilever snaps back to its equilibrium
position, the assumption that the force on the molecule ($F_m$) is
balanced by the cantilever spring restoring force ($F_s$) no longer
holds. Therefore, using the measured $F_s$ gave rise to an overestimate
of the free energy.  
Note that even though the snapping process is
almost instantaneous (small change in $t$, hence $\lambda$), the change
in $z$ is significant because $\Delta z=F_s/k_s$, where $k_s$ is the
cantilever spring constant (see Fig.~\ref{stretching}).
A lower pulling velocity and
larger spring constant will reduce the size of the snapping region.
However, we can estimate the folding free energy barrier, 
$\Delta G_f^\ddagger$, from the equilibrium unfolding 
free energy determined from chemical denaturant studies 
\cite{Baker01a}.  Using $\Delta G_{u}=7.5$ kcal/mol,
we obtained $\Delta G_f^\ddagger=3.9$ kcal/mol, in the expected range 
for titin I27. Figure~\ref{stretching} summarizes the reconstructed free
energy surface and its relation to pulling experiments.

One requirement for using Jarzynski's equality is that the schedule for
varying the work parameter $\lambda$ must be pre-determined
\cite{Jarzynski97a}, which means that constant force ramp is not an
appropriate schedule. A constant $dF_s/dt$ requires force feedback and,
since the measured force $F_s$ fluctuates from one pull to another, the
result is a different schedule of $\lambda$ for each realization. On
the other hand, the dynamic force spectroscopy method commonly used in
AFM pulling of proteins is particularly suitable for such analysis
because the schedule for pulling is pre-determined and remains the same
for all experiments performed at the same velocity.

Using nonequilibrium single-molecule measurements and Jarzynski's
equality, we have reconstructed the free energy surface of both mechanical 
stretching and unfolding of the I27 domain of human cardiac titin.  
Since the profile is an equilibrium property, the reverse of the
free energy surface of stretching is equivalent to that of protein 
folding from an extended state.  The unfolding free energy barrier 
and the prefactor were determined directly from
experimental measurements without having to assume either a two- or a
three- state model, which are major sources of error in the event of
populated intermediate states. In fact, with adequate resolution and
accuracy, an intermediate state should be directly resolvable in the
free energy curve. The topography and the roughness of the folding free
energy landscape can also be determined. Reconstruction of free
energy surfaces directly from experimental data is valuable to
obtain fundamental thermodynamic properties such as the free energy
barrier of unfolding, to understand the mechanical properties of
the molecule, and to compare with theory and simulation results
\cite{Onuchic00a}. With a complete characterization of the free energy
surface of molecular processes, questions such as whether thermal,
chemical, and mechanical unfolding probe the same process may be
resolved. Moreover, since the free energy surface is determined in a
particular environment, how the free energy surface changes with
environmental parameters such as temperature, solution ionic
concentration, and acidity may now be evaluated. Quantification of the
molecular response to external parameters should lead to a better understanding
of molecular behavior in the complex cellular environment.

We thank C.\ Jarzynski, D.\ Thirumulai, K.\ W.\ Plaxco, and S.\ S.\
Plotkin for helpful discussions. We also thank NSF DMR-0505814, 
NIH 1T90DK70121-01, Hamill Innovation Fund, and 
Welch Foundation C-1632 for support.

$^*$To whom correspondence should be addressed.

~Electronic address: chkiang@rice.edu \vspace{-.1in}
\bibliography{jarzynski}

\end{document}